\documentclass[aps,prb,twocolumn,floatfix,showpacs]{revtex4}

\usepackage{graphicx} 
\usepackage{calc} 
\usepackage{amsfonts}  
\usepackage{amsmath} 

\begin{document}

\title{The quantum {\it vs} classical aspects of one dimensional electron-phonon systems revisited by the    renormalization group method}        
 
\author{H. Bakrim and C. Bourbonnais}
\affiliation{Regroupement Qu\'ebecois sur les Mat\'eriaux de Pointe,
  D\'epartement de physique, Universit\'e de Sherbrooke, Sherbrooke,
  Qu\'ebec, Canada, J1K-2R1 }

\date{}

\begin{abstract}
An extension of the  renormalization group method  that includes   the effect of retardation for the  interactions of a fermion gas  is used to  re-examine    the quantum and classical properties of  Peierls-like states in one dimension.  For models of spinless and spin-${1\over 2}$ fermions  interacting with either  intra or intermolecular  phonons the quantum corrections to the Peierls gap  at half-filling are determined at arbitrary phonon frequency. The nature of   quantum-classical transitions is clarified in weak coupling.\end{abstract}
\maketitle 

 \section{Introduction}
The  influence exerted by  zero point ionic motion on the  stability of the Peierls and spin-Peierls lattice distorted states enters as a  key ingredient  in the elaboration  of a general  theoretical description of these phases.   Quantum fluctuations are known to cause a downward renormalization of the order parameter and the corresponding electronic gap, if not  their complete  suppression  as  it is the case  for    spin-Peierls order.   One is confronted to such  situations  in   low dimensional conductors and insulators for which the characteristic phonon energy is not only finite in practice, but  may   exceed  by far the temperature scale at which the lattice instability takes place. These cases are   exemplified  in spin-Peierls systems like the inorganic compound  CuGeO$_3$,\cite{Braden96,Pouget01}   the   organic system     MEM(TCNQ)$_2,$\cite{Pouget01}  and also members of the (TMTTF)$_2$X series  of organic compounds for which  non adiabaticity emerges  as one moves along  the pressure scale, giving rise to quantum criticality for the spin-Peierls transition.\cite{Chow98} 

The first systematic studies of quantum effects on the Peierls-type distorted  states go back in the eighties with the world-line  Monte Carlo simulations of Hirsch and Fradkin.\cite{Fradkin83,Hirsch83} These simulations were made  on the one-dimensional tight binding   and   Holstein electron-phonon models, also  known as the Su-Schrieffer-Heeger (SSH)\cite{Su79}  and  the molecular crystal (MC)\cite{Holstein59} models.    The  stability of  lattice distorted  phases was determined as a function  of the ionic mass and the strength of electron-phonon coupling. The   phase diagrams  of the models were outlined for both spinless and spin-${1\over 2}$ fermions at half-filling.  These initial works were followed by a   variety  of numerical  techniques applied to the same models and  extended  to include direct interactions between fermions.  That  is how   density matrix renormalization group (DMRG),\cite{Caron96,Bursill98,Bursill99}  exact diagonalizations,\cite{BakrimDiag}  and quantum Monte Carlo \cite{BakrimMC} techniques to mention a few, have contributed to  provide  a fairly coherent picture of the  influence  wielded by zero point lattice fluctuations in one-dimensional  electron-phonon systems. 
 
On the analytical side, these progress were preceded\cite{BakrimBefore,Grest76} and accompanied  \cite{Fradkin83,Hirsch83,Caron84,Kuboki87,BakrimDIV1,BakrimDIV2,Uhrig98,Citro05} by    a whole host of approaches  applied to  study retardation effects on lattice distortion at intermediate  phonon frequencies. The     renormalization group  (RG) method\cite{Caron84,BakrimVoit,BakrimRG1,Uhrig98,Citro05}  has been  one of the routes   proposed to deal with  this problem.  A variant of the RG method  will be  further   developed in  this work.  Our analysis starts with the effective fermionic  formulation of the electron-phonon problem, which  is    expressed     in terms of  a  fermion gas in the continuum with weak retarded interactions.  Such a formulation for   the SSH and MC models  has been investigated  long ago  by  the two-cutoff scaling method.\cite{Caron84}   In this approach the characteristic bandwidth energy $E_0$  for fermions and the  vibrational energy $\omega_c$ ($\hbar =1$) for phonons determine the form of flow equations for the electronic scattering amplitudes,\cite{Grest76}    whose singularities  signal   the creation of gaps and  long-range order at half-filling. Thus when the electronic mean-field energy gap $\Delta_0$  $-$ emerging below $E_0$ in the adiabatic weak coupling  theory $-$ is larger than $\omega_c$,   quantum corrections are neglected and   the flow  is equivalent  to a ladder diagrammatic summation  compatible with  the unrenormalized static  scale  $\Delta_0$ for the gap.  On the other hand when $\Delta_0 < \omega_c$,   the scattering amplitudes, though  still governed by the ladder flow down to $\omega_c$, are     considered as effective  unretarded interactions at lower energies. Below  $\omega_c$ the flow becomes impregnated by vertex corrections and   interference between different scattering channels. In accord with  the well known results of the one-dimensional  electron gas model, \cite{Emery79,Solyom79,Kimura75,Bourbon91} the classical gap $\Delta_0$ is then an irrelevant scale and the system  enters in the non adiabatic quantum  domain where  either a gapless  or an ordered massive phase  can occur. 
 
While  the    two-cutoff RG analysis can provide  a  simple and reliable criteria to  map  out the essentials of the  quantum-classical boundaries of the phase diagram for  both models in the weak coupling sector, \cite{Caron96,Bursill98} it says nothing, on the other hand, on how the gap  varies over the whole phonon frequency range. This  is not only of practical importance, when   {\it e.g.,} the  theory is confronted  to experiment in concrete cases,   but also clearly needed on general grounds when  one raises the question  of the nature of quantum-classical transition as a function of  phonon frequency. This drawback is not a weakness of the RG method in general but  rather ensues from the frequency dependence of couplings, which in the two cut-off scaling approach, barely  reduces  to  the minimum  found in either the  adiabatic or  non adiabatic limit. A continuous description of retardation  effects would require    that the full functional dependence of scattering amplitudes  on the frequencies  be restored, a possibility that  can be liken to what  has been  done in two-dimensional and  quasi-one-dimensional RG  for the functional dependence on scattering amplitudes on the momentum.\cite{Zanchi00,Honerkamp00,Halborth00}$^,$ \cite{Nickel06,Tsuchiizu06,Duprat01} Very recent progress along these lines show that it is indeed  a promising avenue. \cite{Bakrim1}

In this paper we shall revert to the  RG approach as developed in Refs. \cite{Bourbon91,Bourbon03} and extend    its   formulation  to  include  the frequency dependence of scattering amplitudes   introduced by the electron-phonon interaction. We revisit the classical and quantum aspects of fermion driven lattice  instabilities.  Our  analysis is done at the one-loop level  and covers the gap determination and the structure of the phase diagram of the MC and SSH models   for both spinless and spin-${1\over 2}$ fermions. Although the  generalization to   incommensurate band filling   and situations where the direct   Coulomb interaction is included would cause  no difficulty, we have  restricted  our analysis  to retarded interactions    at  half-filling.  In Sec.~II, we introduce the electron-phonon models and recall the derivation of their respective bare  retarded interactions in the framework of an effective fermion gas model.  We pay special attention  to   the SSH model in the spinless case in order to include the momentum dependent  umklapp term to the interaction parameter space, which is so important for long-range order of  this model. In Sec.~III  the one-loop level  flow equations for the retarded   scattering amplitudes and  response functions   are derived for   spinless and spin-${1\over 2}$ fermions. In Sec.~IV we compute the variations of the gap over the  whole frequency range and  discuss the structure of the phase  diagram and the nature of the quantum-classical transitions for the MC and SSH models. We conclude in Sec.~V. 
\section{The models and the partition function}
\label{mod}
\subsection{Models}
The   one-dimensional electron-phonon models that  we shall  study using the RG  method are the   MC  and the  SSH  models. The MC model  describes the coupling of fermions to optical  molecular phonon modes, whereas for the SSH model the electron-phonon interaction results from the modulation of electronic energy by acoustic phonons.   In Fourier space, the two one-dimensional models Hamiltonians can be  written in  following   form
\begin{eqnarray}
H &= & H_{0} + H_{\rm ph} + H_{\rm I}\cr
& = &\sum_{k,\sigma} \epsilon(k) c^\dagger_{k,\sigma}c_{k,\sigma} + \sum_q \omega_q\Big(b^\dagger_qb_q + {1\over2}\Big)  \cr  
&+ & L^{-{1\over 2}}\sum_{k,q,\sigma} g(k,q)c^\dagger_{k + q,\sigma}c_{k,\sigma}(b^\dagger_q + b_{-q}). 
\label{GHamiltonian}
\end{eqnarray}
Here  $H_{0} $ is the free fermion part   and $\epsilon(k)=-2t \cos k$  is the tight-binding energy spectrum  with $t$ as the hopping integral (the lattice constant $a=1$ and $L$ is the number of sites). $c^\dagger_{k,\sigma}$ $(c_{k,\sigma})$ creates (annihilates) a  fermion of wave vector $k$ and spin $\sigma$.   $ H_{\rm ph}$ and $ H_{\rm I}$ terms correspond  to  the free phonon and electron-phonon interaction  parts, respectively,  and in which $b^\dagger_{q}$ $(b_{q})$ creates (annihilates) a  phonon of wave vector $q$. For the MC model,\cite{Holstein59}  the   intramolecular phonon energy  and  the  interaction are given by 
\begin{eqnarray}
\label{MC}
\omega_q& = &\omega_0, \\
g(k,q) & = &  {\lambda_0/ \sqrt{2M_0\omega_0}},
\label{MCg}
\end{eqnarray}
which  are both independent of the momentum.   Here $\lambda_0 >0$ is the amplitude of the electron-phonon interaction on each molecular site whereas   
 the frequency $\omega_0=\sqrt{\kappa_0/M_0} $ is expressed in terms of the elastic constant $\kappa_0$ and the molecular mass $M_0$.  
 
 For the SSH model,\cite{Su79} the corresponding quantities read
\begin{eqnarray}
\omega_q & = & \omega_D\big|\sin{q\over 2}\big|, \\
g(k,q) & = & i 4{\lambda_D\over \sqrt{2M_D \omega_D}} \sin{q\over 2} \cos\big(k + {q\over 2}\big),
\label{SSH}
\end{eqnarray} 
where $\omega_D =2\sqrt{\kappa_D/M_D}$ is  the acoustic  phonon energy at $q=2k_F$, namely at  twice the Fermi wave vector $k_F=\pi/2$ at half-filling. $M_D$ is the ionic mass and $\kappa_D$ is the constant force of the one-dimensional lattice.

 \subsection{The partition function}
 \label{Z}
Following  the trace over  harmonic phonon degrees of freedom in the interaction Matsubara time representation of the grand canonical partition function $Z$, one can write   
\begin{widetext}
\begin{eqnarray}
\label{}
Z & = & {\rm Tr}_{\rm e} \, e^{-\beta H_0-\mu N} \, {\rm Tr}_{\rm ph}e^{-\beta H_{\rm ph}}\,  T_\tau \exp \bigg\{\!-\!\!\int_0^\beta H_{\rm I}(\tau)d\tau \bigg\} \cr 
&=& Z_{\rm ph}{\rm Tr}_{\rm e} \, e^{-\beta H_0-\mu N}  T_\tau \exp \bigg\{\!-{1\over 2}\!\!\sum_{\{k,q,\sigma\}}\!\int_0^\beta \!\!\int_0^\beta g(k,q)g(k',-q)D(q,\tau-\tau')   c^\dagger_{k + q,\sigma}(\tau)c^\dagger_{k'- q,\sigma'}(\tau')c_{k',\sigma'}(\tau')c_{k,\sigma}(\tau) d\tau'd\tau \bigg\}\cr &&  \end{eqnarray}
\end{widetext}
where $Z_{\rm ph}$ is the partition function of bare phonons. The phonon integration   introduces an effective `retarded' fermion interaction mediated by phonons and described by the bare  propagator 
$$ 
D(q,\tau-\tau')= e^{-\omega_q|\tau-\tau' |}
+2(e^{\beta\omega_q}-1)^{-1} \cosh\bigl(\omega_q(\tau -\tau')\bigr).
$$
The remaining trace over fermion degrees of freedom  can be recast into a functional integral form  
\begin{eqnarray}
\label{Z}
Z &= &Z_{\rm ph}  \int\!\!\!\int \!\mathfrak{D}\psi^*\mathfrak{D}\psi\ e^{S[\psi^*,\psi]},\cr
 &=&Z_{\rm ph}  \int\!\!\!\int \!\mathfrak{D}\psi^*\mathfrak{D}\psi\ e^{S_0[\psi^*,\psi] + S_I[\psi^*,\psi]},
 \end{eqnarray}
over  the anticommuting  Grassman fields $\psi$. In the Fourier Matsubara space, the free fermionic action is    
\begin{equation}
\label{ }
S_0[\psi^*,\psi] = \sum_{p,\tilde{k},\sigma} [G^0_p(\tilde{k})]^{-1} \psi^*_{p,\sigma}(\tilde{k}) \psi_{p,\sigma}(\tilde{k}),
\end{equation}
where
\begin{equation}
\label{ }
G^0_p(\tilde{k}) =\big[i\omega -\epsilon_p(k)\big]^{-1}
\end{equation}
is the bare fermion propagator for \hbox{$\tilde{k}= ( k,\omega =\pm \pi T,\pm 3\pi T, \ldots)$} ($k_B=1$).   The fermion spectrum \hbox{$\epsilon(k)-\mu \approx \epsilon_p(k)= v_F(pk-k_F)$} is linearized around the  right $(p=+)$ and left $(p=-$) Fermi points $\pm k_F$. The bandwidth  cut-off $E_0=2E_F$ is  twice  the Fermi energy \hbox{$  E_F=  v_Fk_F$}. The integration of the fermion degrees of freedom becomes \hbox{$\int\!\!\int \mathfrak{D}\psi^*\mathfrak{D}\psi =\int\!\!\int \prod_{p,\sigma,\tilde{k}} d\psi^*_{p\sigma}(\tilde{k}) d\psi_{p\sigma}(\tilde{k})$}.

 The interacting part $S_I$ of the action reads
\begin{widetext}
\begin{eqnarray}
S_I[\psi^*,\psi]  &=& -{T\over 2 L} \sum_{\{p,\tilde{k},\sigma\}}g( \tilde{k}_1,\tilde{k}_2; \tilde{k}_3,\tilde{k}_4) \psi^*_{p_1,\sigma_1}(\tilde{k}_1)\psi^*_{p_2,\sigma_2}(\tilde{k}_2) \psi_{p_4,\sigma_2}(\tilde{k}_4) \psi_{p_3,\sigma_1}(\tilde{k}_3) \delta_{k_{1+2},k_{3+4} +G }\delta_{\omega_{1+2},\omega_{3+4}},\cr
&& \ 
\end{eqnarray}
\end{widetext}
where momentum conservation is assured modulo the reciprocal lattice vector $G=\pm4k_F$, allowing for umklapp scattering at half-filling. In the Fourier-Matsubara space, the interaction takes the form   
\begin{equation}
g( \tilde{k}_1,\tilde{k}_2; \tilde{k}_3,\tilde{k}_4) =    g(k_1,k_3-k_1 ) g(k_2,k_4-k_2)D(\tilde{k}_3-\tilde{k}_1),
\label{coupling}
\end{equation}
where 
$$
D(\tilde{k}_3-\tilde{k}_1)= -2{\omega_{k_3-k_1}\over \omega_{k_3-k_1}^2 + \omega_{3-1}^2},
$$
is the   bare phonon propagator. We can now proceed to the `g-ology' decomposition of this interaction. This will be done  separately for fermions with and without spins. 

In the first place, for  spin-${1\over 2}$ fermions, we shall consider    the three standard  couplings between fermions on opposite Fermi points
 \begin{eqnarray}
 g_1(\omega_1,\omega_2,\omega_3)  &\equiv & g (\pm k_F,\omega_1,\mp k_F,\omega_2;\mp k_F,\omega_3,\pm k_F,\omega_4),  \cr\cr
    g_2(\omega_1,\omega_2,\omega_3) &\equiv  & g (\pm k_F,\omega_1,\mp k_F,\omega_2; \pm k_F,\omega_3,\mp k_F,\omega_4), \cr\cr    g_3(\omega_1,\omega_2,\omega_3) & \equiv & g (\pm k_F,\omega_1,\pm k_F,\omega_2; \mp k_F, \omega_3,\mp k_F,\omega_4),\cr &&\    
\end{eqnarray}
for  retarded backward, forward and umklapp scattering amplitudes, respectively  (here the forward scattering of fermions on the same branch is neglected). According to Eq.~(\ref{MCg}), the bare frequency dependent couplings for the MC model become  
\begin{eqnarray}
 g_{i=1,2,3}(\omega_1,\omega_2,\omega_3)   =       { g_i\over 1 + {\omega_{3-1}^2/ \omega_0^2}},
 \label{IniMC}   
\end{eqnarray}
where $g_{i=1,2,3}= -\lambda_0^2/\kappa_0 $ is the ($M_0$-independent) attractive amplitude. Similarly for the SSH model,  one has
from Eq.~(\ref{SSH})\begin{eqnarray}
g_{1,3}(\omega_1,\omega_2,\omega_3) & = &  { g_{1,3} \over 1 + {\omega_{3-1}^2/ \omega_D^2}},
\label{IniSSH}
\end{eqnarray}
where the amplitudes  $g_{1,3} =   \mp4\lambda^2_D/ \kappa_D $ are also $M_D$-independent. For the SSH model, the bare forward scattering amplitude $g_2$ vanishes for the exchange of zero momentum phonon but it will be generated at lower energy  by the renormalization group transformation.   

For spinless fermions, the backward scattering is   indistinguishable by exchange from  the forward scattering  and both can be  combined to define  an effective forward scattering term of the form 
\begin{eqnarray}
g_f(\omega_1,\omega_2,\omega_3) & \equiv &g(\pm k_F,\omega_1,\mp k_F,\omega_2; \pm k_F,\omega_3,\mp k_F,\omega_4)\cr
& -&   g (\pm k_F,\omega_2,\mp k_F,\omega_1;\mp k_F,\omega_3,\pm k_F,\omega_4)\cr
&=&    { g_2\over 1 + {\omega_{3-1}^2/ \omega_{0,D}^2}} -{g_1\over 1 + {\omega_{3-2}^2/  \omega_{0,D}^2} },
\label{gf} 
 \end{eqnarray}
where the mass independent amplitudes are \hbox{$ g_{1,2}=-\lambda_0^2/\kappa_0$}  for the MC model,  and  \hbox{$g_{1}= -4\lambda_D^2/\kappa_D  $} and $g_{2} =0$ in the  SSH case. 

As for the umklapp scattering in the spinless case, it  must be antisymmetrized with   its own exchange term to give  the following two contributions 
$$
 {1\over 2} [  g( \tilde{k}_1,\tilde{k}_2; \tilde{k}_3,\tilde{k}_4)   -   g( \tilde{k}_2,\tilde{k}_1; \tilde{k}_3,\tilde{k}_4)]^{k_{3}\sim k_4}_{k_{1}\sim k_2}\equiv    \,g_3(\omega_1,\omega_2,\omega_3)  
 $$
\begin{eqnarray}
  +\    g_u(\omega_1,\omega_2,\omega_3) 
(\sin k_1-\sin k_2)  (\sin k_3-\sin k_4). 
   \label{g3S0}  
  \end{eqnarray}
 The first contribution corresponds to a local umklapp term defined for incoming and outgoing fermions {\it at}  the Fermi points. It takes the form  
\begin{eqnarray}
\label{g3local}
 g_3(\omega_1,\omega_2,\omega_3)\!\! & = & \!\! {1\over 2}\,[g(\pm k_F,\omega_1,\pm k_F,\omega_2; \mp k_F,\omega_3,\mp k_F,\omega_4)\cr
  & - &  g(\pm k_F,\omega_2,\pm k_F,\omega_1; \mp k_F,\omega_3,\mp k_F,\omega_4)]\cr
  &=& { {g_3 }\over 1 + {\omega_{3-1}^2/ \omega_{0,D}^2}} -{{g_3 }\over 1 + {\omega_{3-2}^2/ \omega_{0,D}^2}},
\end{eqnarray}
This term is present for both models, where  \hbox{$g_3= -\lambda_0^2/\kappa_0 $} is attractive for the MC model, and \hbox{$g_3=   4\lambda_D^2/\kappa_D$} is repulsive for the SSH model.   
 The second term of (\ref{g3S0}) is a non local -- momentum dependent -- umklapp contribution  and is only present for the SSH model. Actually  this additional  contribution  follows from the antisymmetrization of Eq.~(\ref{coupling}) and the use of  (\ref{SSH})  under the permutation of incoming and outgoing frequencies and  momentum (these last,   {\it not} at the Fermi points). Its frequency dependent part reads  
 \begin{equation}
\label{guk}
g_u(\omega_1,\omega_2,\omega_3)= g_u \Big[{ 1\over 1 + {\omega_{3-1}^2/ \omega_D^2}} +{1\over 1 + {\omega_{3-2}^2/ \omega_D^2}}\Big],
\end{equation}
where the amplitude is given by  $g_u=  \lambda^2_D/2\kappa_D$.
From Eq.~(\ref{g3S0}), it follows that  for $k_{1(3)}\sim k_{2(4)}$, the leading $k$-dependence of the non local umklapp     is $\propto (k_1-k_2)(k_3-k_4) $ which has a  scaling dimension of -2.  This term  is therefore  strongly irrelevant at the tree level  but becomes  relevant  beyond some  threshold in the electron-phonon interaction. Such umklapp contributions  are well known to play   a key role in  the  existence  of long-range order  for interacting spinless fermions, \cite{denNijs81,Black81}  as it will  show to be the case   for the SSH model.\cite{Fradkin83,Caron84,Caron96}  
\section{The renormalization group transformation}
The renormalization group transformation for the partition function will follow  the one given in Ref.\cite{Bourbon91,Bourbon03}  One then proceeds for $Z$ to the successive  partial integration of fermion degrees of freedom, denoted by $ \bar{\psi}^{(*)}$ having the momentum located in the outer energy shells (o.s) $\pm E_0(\ell)d\ell/2$ above and below the Fermi points for each fermion branch $p$. The remaining $(<)$ degrees of freedom are kept fixed.  Here $E_0(\ell)=E_0e^{-\ell}$ is the scaled bandwidth at   $\ell\ge 0$.  The integration proceeds by first splitting the action \hbox{$S\to S[\psi^*,\psi]_\ell + \bar{S}_0 +\bar{S}_I$} into an inner   shell part at $\ell$ and    the  \hbox{$ \bar{\psi}$ --} dependent outer shell terms $\bar{S}_0$ and $\bar{S}_I$. Considering   $\bar{S}_I$ as a perturbation  with respect to  the free outer-shell action $\bar{S}_0$, the partial integration  at the one-loop level  is of the form
\begin{eqnarray}
\label{ZRG}
Z  
   \! & \sim  &\!\! \int\!\!\int_<  \mathfrak{D}\psi^*   \mathfrak{D}\psi \ {  e}^{S[\psi^*,\psi]_\ell  } \cr
  &&\ \ \  \times  \int\!\!\int_{\rm o.s}  \ \mathfrak{D}\bar{\psi}^* \mathfrak{D}\bar{\psi} \ e^{\bar{S}_0[\bar{\psi}^*,\bar{\psi}] + \bar{S}_I[\bar{\psi}^*,\bar{\psi},\psi^*,\psi] }  \cr
   & \propto & \!\!\int\!\!\int_< \mathfrak{D}\psi^*   \mathfrak{D}\psi  \ e^{S[\psi^*,\psi]_\ell +  \langle \,\bar{S}_{I} \,\rangle_{\rm o.s}+  {1\over2} \langle \,\bar{S}_{I} ^2\,\rangle_{\rm o.s} +   \, \ldots}. 
   \label{Kadanoff}
     \end{eqnarray}
Here  the  interacting part is  made up  of three pertinent terms, {\it i.e.,} \hbox{$\bar{S}_I= \bar{S}_{I,2}^P + \bar{S}_{I,2}^C +\bar{S}_{I,2}^L$}, for all possibilities of putting simultaneously two outer shell fields in the  $2k_F$ electron-hole  Peierls channel   (\hbox{$\bar{S}_{I,2}^P \sim \bar{\psi}^*_{+}{\psi}^*_{-}\bar{\psi}_-\psi_+ $}  +  \hbox{$ \bar{\psi}^*_{+}{\psi}^*_{+}\bar{\psi}_-\psi_- +\ldots $}), the zero momentum fermion-fermion Cooper channel (\hbox{$ \bar{S}_{I,2}^C \sim \bar{\psi}^*_{+}\bar{\psi}^*_{-}\psi_-\psi_+ +\ldots)   $}, and the Landau channel  \hbox{$ (\bar{S}_{I,2}^L\sim   {\psi}^*_{+}\bar{\psi}^*_{-}\bar{\psi}_-\psi_+  + \ldots)$}.

The lowest order  outer shell statistical average  $\langle \,\bar{S}_{I} \,\rangle_{\rm o.s}$ comes from  the Landau part and gives rise to   the   self-energy corrections $\delta\Sigma(\omega)$ of the one-particle Green function, which becomes $G_p^{-1} + i\delta \Sigma(\omega)$. As for the contractions ${1\over 2}\langle \,\bar{S}_{I} ^2\,\rangle_{\rm o.s}$, only the  singular Peierls and Cooper scattering channels are retained; with four fields in the inner shell, these  correspond to   corrections to the coupling constants. Both corrections   define the renormalized action $S[\psi^*,\psi]_{\ell+d\ell}$ at the  step $\ell+d\ell$.

The evaluation of outer shell contractions  $\langle \,\bar{S}_{I} \,\rangle_{\rm o.s}$ for the self-energy at the one-loop level leads to 
\begin{eqnarray}
 \langle \,\bar{S}_{I} \,\rangle_{\rm o.s}  & = & i \, \delta\Sigma (\omega)\sum_{p,\tilde{k}_<,\sigma}  \psi^*_{p,\sigma}(\tilde{k})  \psi_{p,\sigma}(\tilde{k}), \cr
\delta\Sigma (\omega) & = & -\pi v_F{ T\over L} \sum_{\omega'}\!\!\sum_{\ \ \{k\}_{\rm o.s}} \tilde{g}_s(\omega',\omega,\omega)G_-(k ,\omega')  .\cr
&&
\end{eqnarray}
 In the low temperature limit, the flow equation  for the self-energy becomes 
\begin{eqnarray}
\partial_\ell   \Sigma(\omega) =  \int_{-\infty}^{+\infty} {d\omega'\over 2\pi}\!\!\!\!\!&&\!\!\!\Big\{  \tilde{g}_s(\omega',\omega,\omega) \cr
&\times&
\frac{(E_{0}(\ell)/2) (\omega'-\Sigma (\omega') )}{ (\omega'-\Sigma_{p} (\omega'))^{2}+ (E_{0}  (\ell)/2 )^{2}}\Big\},\cr
&&
\label{Flself}
 \end{eqnarray}
where
\begin{equation}
\label{gself}
\tilde{g}_s(\omega',\omega,\omega) = \tilde{g}_{1}(\omega',\omega,\omega)-2\tilde{g}_{2}(\omega,\omega',\omega),
\end{equation} 
for spin-${1\over 2}$ fermions, and
\begin{equation}
\label{gZC}
\tilde{g}_s(\omega',\omega,\omega) = -\tilde{g}_{f}(\omega,\omega',\omega),
\end{equation}
in the spinless case. Here   $\tilde{g}_i(\{\omega\})\equiv g_i(\{\omega\})/\pi v_F$, $\partial_\ell\equiv \partial/\partial\ell $, and $ \Sigma (\omega)=0$ at $\ell=0$. 
\subsection{RG flow for couplings:  spin-${1\over 2}$ fermions}

The one-loop contractions ${1\over 2}\langle \,\bar{S}_{I} ^2\,\rangle_{\rm o.s}$ amount to evaluate the outer shell contributions of the Peierls [${1\over 2} \langle (S_{I,2}^P)^2\rangle_{\rm o.s.} ]$ and Cooper  [${1\over 2} \langle (S_{I,2}^C)^2\rangle_{\rm o.s.} ]$ channels. For the MC and SSH models defined by the  initial couplings (\ref{IniMC}-\ref{IniSSH}), these interfering contractions lead  to the   flow equations  \begin{widetext}
\begin{eqnarray}
\label{Flowg1}
\partial_\ell \tilde{g}_{1}(\omega_1,\omega_2,\omega_3) = \int_{-\infty}^{+\infty}{d\omega\over 2\pi}   \Big\{  &-& 2\,\tilde{g}_1(\omega,\omega_2,\omega+\omega_3-\omega_1)\tilde{g}_1(\omega_1,\omega+\omega_3-\omega_1,\omega_3) I_P(\omega,\omega_3-\omega_1) \cr
& + &  \tilde{g}_1(\omega,\omega_2,\omega+\omega_3-\omega_1)\tilde{g}_2(\omega+\omega_3-\omega_1,\omega_1,\omega_3)  I_P(\omega,\omega_3-\omega_1) \cr\cr & + & \tilde{g}_1(\omega_1,\omega,\omega_3)\tilde{g}_2(\omega_2, \omega+\omega_1-\omega_3,\omega) I_P(\omega,\omega_1-\omega_3)\cr\cr
& - &  2\, \tilde{g}_3(\omega_2,\omega,\omega_1+\omega_2-\omega_3)\tilde{g}_3(\omega_1,\omega+\omega_3-\omega_1,\omega_3) I_P(\omega,\omega_3-\omega_1)\cr\cr
& + &  \tilde{g}_3(\omega_1,\omega,\omega_3)\tilde{g}_3(\omega+\omega_1-\omega_3,\omega_2,\omega_1+\omega_2-\omega_3)I_P(\omega,\omega_1-\omega_3)\cr\cr
& + &   \tilde{g}_3(\omega_2,\omega,\omega_1+\omega_2-\omega_3)\tilde{g}_3(\omega+\omega_3-\omega_1,\omega_1,\omega_3) I_P(\omega,\omega_3-\omega_1)\cr\cr
 &-& \Big[\  \tilde{g}_2(\omega_2,\omega_1,\omega+\omega_1+\omega_2)\tilde{g}_1(-\omega,\omega_2,\omega_1+\omega_2+\omega)\cr\cr
& + & \tilde{g}_1(\omega_1,\omega_2,\omega+\omega_1+\omega_2)\tilde{g}_2(\omega+\omega_1+\omega_2,-\omega,  \omega_3) \Big]I_C(\omega,\omega_1+\omega_2)\Big\}\cr
&&\\
\label{Flowg2}
\partial_\ell \tilde{g}_{2}(\omega_1,\omega_2,\omega_3) = \int_{-\infty}^{+\infty}{d\omega\over 2\pi}  
 \Big\{  & \Big[& \tilde{g}_2(\omega_1,\omega+\omega_2-\omega_3,\omega)\tilde{g}_2(\omega,\omega_2,\omega_3) \cr &+ &\tilde{g}_3(\omega_1,\omega+\omega_2-\omega_3,\omega)\tilde{g}_3(\omega,\omega_2,\omega_3) \Big] I_P(\omega,\omega_2-\omega_3) \cr\cr
&-&\Big[\  \tilde{g}_2(\omega_1,\omega_2,\omega+\omega_1+\omega_2)\tilde{g}_2(  \omega+\omega_1+\omega_2,  -\omega, \omega_3)\cr\cr
                     &+& \tilde{g}_1(\omega_2,\omega_1,\omega+\omega_1+\omega_2)\tilde{g}_1(-\omega,\omega+\omega_1+\omega_2,\omega_3)\Big]I_C(\omega,\omega_1+\omega_2) \Big\} \cr
                     &&\\
                      \label{Flowg3}
 \partial_\ell \tilde{g}_{3}(\omega_1,\omega_2,\omega_3) = 2 \int_{-\infty}^{+\infty}{d\omega\over 2\pi}  
 \Big\{   &- & 2\, \tilde{g}_1(\omega_1,\omega+\omega_3-\omega_1,\omega_3) \tilde{g}_3(\omega,\omega_2,\omega+\omega_3-\omega_1)I_P(\omega,\omega_3-\omega_1) \cr
    &  +&   \tilde{g}_1(\omega_1,\omega,\omega_3) \tilde{g}_3(\omega+\omega_1-\omega_3,\omega_2,\omega_1+\omega_2-\omega_3)I_P(\omega,\omega_1-\omega_3)\cr\cr
    &+&   \tilde{g}_3(\omega_1,\omega,\omega_3)\tilde{g}_2(\omega+\omega_1-\omega_3,\omega_2,\omega_1+\omega_2-\omega_3)  I_P(\omega,\omega_1-\omega_3) \cr\cr  
    &+& \tilde{g}_3(\omega,\omega_2,\omega_3 )\tilde{g}_2(\omega+\omega_2-\omega_3,\omega_1, \omega_1+\omega_2-\omega_3) I_P(\omega,\omega_2-\omega_3)\Big\}.\cr           
&& \ 
 \end{eqnarray}
 \end{widetext}
 The momentum shell     Peierls and Cooper loops  $I_{P}(\omega,\Omega) $ and  $I_{C}(\omega,\Omega) $ at internal $\omega$ and their respective external frequency  $\Omega$ are by using the Green function with self-energy corrections 
\begin{widetext}
\begin{eqnarray}
 \label{loopP}
I_P(\omega,\Omega)d\ell & = &-  {\pi v_F\over L} \sum_{\ \ \{k\}_{\rm o.s}} G_+(k+2k_F,\omega +\Omega)G_-(k ,\omega), \cr
                        & = &{d\ell\over 2} E_0(\ell) {(\omega-\Sigma (\omega))(\omega+\Omega-\Sigma (\omega+\Omega))  + {1\over4}E_0^2(\ell)\over [(\omega-\Sigma (\omega))^2  +{1\over4}E_0^2(\ell)][(\omega+\Omega-\Sigma (\omega+\Omega))^2 + {1\over4}E_0^2(\ell)]},
\end{eqnarray}
\begin{eqnarray}
 \label{loopC}
I_C(\omega,\Omega)d\ell & = &  {\pi v_F\over L} \sum_{\ \ \{k\}_{\rm o.s}} G_+(k,\omega +\Omega)G_-(-k,-\omega),\cr
                        & = &{d\ell\over 2} E_0(\ell) {(\omega-\Sigma (\omega))(\omega-\Omega+\Sigma (\Omega-\omega))  + {1\over4}E_0^2(\ell)\over [(\omega-\Sigma (\omega))^2  +{1\over4}E_0^2(\ell)][(\omega-\Omega+\Sigma (\Omega-\omega))^2 + {1\over4}E_0^2(\ell)]}.
\end{eqnarray}

\end{widetext}

\subsection{RG flow for couplings: spinless fermions \label{RGspinless}}
Owing to the nature of umklapp scattering which  in the spinless case is different for the MC and SSH models, we shall proceed separately for each model.  Thus for the MC model with a local  umklapp term, the outer shell  contractions  $ {1\over 2} \langle (S_{I,2}^P)^2\rangle_{\rm o.s.}$  and $ {1\over 2} \langle (S_{I,2}^C)^2\rangle_{\rm o.s.}$ for the Peierls and Cooper channels allow us to write 
\begin{widetext}
\begin{eqnarray}
\partial_\ell \tilde{g}_{f}(\omega_1,\omega_2,\omega_3) = \int_{-\infty}^{+\infty}{d\omega\over 2\pi}  
 \Big\{  & \Big[& \tilde{g}_f(\omega_1,\omega+\omega_2-\omega_3,\omega)\tilde{g}_f(\omega,\omega_2,\omega_3) \cr &+ &\tilde{g}_3(\omega_1,\omega+\omega_2-\omega_3,\omega)\tilde{g}_3(\omega,\omega_2,\omega_3) \Big] I_P(\omega,\omega_2-\omega_3) \cr\cr
&-&\  \tilde{g}_f(\omega_1,\omega_2,\omega+\omega_1+\omega_2)\tilde{g}_f(  \omega+\omega_1+\omega_2,  -\omega, \omega_3)I_C(\omega,\omega_1+\omega_2) \Big\},\cr
&&\ 
 \label{FlMCgfS0}
\end{eqnarray}
\begin{eqnarray}
\label{FlMCg3S0}
\partial_\ell \tilde{g}_{3}(\omega_1,\omega_2,\omega_3) = 2 \int_{-\infty}^{+\infty}{d\omega\over 2\pi}  
 &\Big\{&      \tilde{g}_3(\omega_1,\omega,\omega_3)\tilde{g}_f(\omega+\omega_1-\omega_3,\omega_2,\omega_1+\omega_2-\omega_3)  I_P(\omega,\omega_1-\omega_3) \cr\cr  
    &+& \tilde{g}_3(\omega,\omega_2,\omega_3 )\tilde{g}_f(\omega+\omega_2-\omega_3,\omega_1, \omega_1+\omega_2-\omega_3) I_P(\omega,\omega_2-\omega_3)\Big\},
    \cr
    && \ 
\end{eqnarray}

\end{widetext}
which are subjected to the initial   conditions (\ref{gf}) and (\ref{g3local}) of the MC model.

The flow equations for  the SSH model presents an important difference  because of the additional   $k-$dependent umklapp term   (\ref{guk}). At variance with $g_f$ and $g_3$, this coupling   acquires a non zero  scaling dimension   at the tree level. Therefore the momentum, energies and fields must be rescaled after each partial trace operation  in  Eq.~(\ref{Kadanoff}), which restores the original bandwidth  cutoff.  Thus following the outer shell integration, one applies  the transformations \hbox{$k'=sk$, $\omega'=s\omega$, $\psi^{(*)'}=s^{-1/2}\psi^{(*)}$, $T'=sT$, $L'=s^{-1}L$}, and $\omega_D'= s\omega_D$ ($M_D'=s^{-2}M_D$, the spring constant $\kappa_D$ is kept fixed),  where $s=e^{d\ell}$. This gives   the scaling transformations for the outer shell corrected couplings, namely $\tilde{g}_{f,3}^\prime(\{\omega'\})= s^0\tilde{g}_{f,3}(\{\omega\})$  for the  local part   and $\tilde{g}_{u}^\prime(\{\omega'\})= s^{-2}\tilde{g}_{u}(\{\omega\})$ for the non local part. The flow equations for the SSH model then become 
\begin{widetext}
\begin{eqnarray}
\partial_\ell \tilde{g}_{f}(\omega_1,\omega_2,\omega_3) = \int_{-\infty}^{+\infty}{d\omega\over 2\pi}  
 \Big\{  & \Big[& \tilde{g}_f(\omega_1,\omega+\omega_2-\omega_3,\omega)\tilde{g}_f(\omega,\omega_2,\omega_3) +\tilde{g}_3(\omega_1,\omega+\omega_2-\omega_3,\omega)\tilde{g}_3(\omega,\omega_2,\omega_3)  \cr\cr
&+ &\tilde{g}_u(\omega_1,\omega+\omega_2-\omega_3,\omega)\tilde{g}_3(\omega,\omega_2,\omega_3)  +\tilde{g}_3(\omega_1,\omega+\omega_2-\omega_3,\omega)\tilde{g}_u(\omega,\omega_2,\omega_3)  \cr\cr
&+ &\tilde{g}_u(\omega_1,\omega+\omega_2-\omega_3,\omega)\tilde{g}_u(\omega,\omega_2,\omega_3) \Big] I_P(\omega,\omega_2-\omega_3) \cr\cr
&-&\  \tilde{g}_f(\omega_1,\omega_2,\omega+\omega_1+\omega_2)\tilde{g}_f(  \omega+\omega_1+\omega_2,  -\omega, \omega_3)I_C(\omega,\omega_1+\omega_2) \Big\},
 \label{gfSSHS0}
\end{eqnarray}
\begin{eqnarray}
\partial_\ell \tilde{g}_{3}(\omega_1,\omega_2,\omega_3) = 2 \int_{-\infty}^{+\infty}{d\omega\over 2\pi}  
 &\Big\{&      \tilde{g}_3(\omega_1,\omega,\omega_3)\tilde{g}_f(\omega+\omega_1-\omega_3,\omega_2,\omega_1+\omega_2-\omega_3)  I_P(\omega,\omega_1-\omega_3) \cr\cr  
    &+& \tilde{g}_3(\omega,\omega_2,\omega_3 )\tilde{g}_f(\omega+\omega_2-\omega_3,\omega_1, \omega_1+\omega_2-\omega_3) I_P(\omega,\omega_2-\omega_3)\Big\},\cr
    && 
\label{g3SSHS0}
\end{eqnarray}
\begin{eqnarray}
\label{guSSHS0}
\partial_\ell \tilde{g}_u(\omega_1,\omega_2,\omega_3)   = &-&2\tilde{g}_u(\omega_1,\omega_2,\omega_3)\cr
&+& 2 \int_{-\infty}^{+\infty}{d\omega\over 2\pi}  
 \Big\{      \tilde{g}_u(\omega_1,\omega,\omega_3)\tilde{g}_f(\omega+\omega_1-\omega_3,\omega_2,\omega_1+\omega_2-\omega_3)  I_P(\omega,\omega_1-\omega_3) \cr\cr  
    &+& \tilde{g}_u(\omega,\omega_2,\omega_3 )\tilde{g}_f(\omega+\omega_2-\omega_3,\omega_1, \omega_1+\omega_2-\omega_3) I_P(\omega,\omega_2-\omega_3)\Big\},
\end{eqnarray}
\end{widetext}
which are subjected to the initial conditions (\ref{gf}), (\ref{g3local}), and (\ref{guk}).

\subsection{Response functions}
Staggered $2k_F$ density-wave and zero pair momentum superconducting susceptibilities  can be  computed by adding a set of linear couplings to composite fields $\{h_\mu\}$ in the bare action   at $\ell=0$. This gives the source field term
\begin{eqnarray}
\label{sources}
 S_h[\psi^*,\psi]   =   \sum_{\omega,\Omega}\Big[\sum_{\mu_{p},M} & \!\!h^M_{\mu_p}(\Omega) z^M_{\mu_p}(\omega,\omega+\Omega)   {\cal O}^{M*}_{\mu_p}(\omega,\Omega)   &\cr 
+ \sum_{\mu_{c}} & \!\! h_{\mu_c}(\Omega) z_{\mu_c}(-\omega,\omega+\Omega)   {\cal O}^*_{\mu_c}(\omega,\Omega) &\cr
& + \ {\rm c.c}\  \Big],&
\end{eqnarray}
where $z^M_{\mu_p}$ and $z_{\mu_c}$ are the renormalization factors  of   the  corresponding source fields,  with $z^{(M)}_{\mu_{c(p)}}=1$ for the boundary conditions at $\ell=0$. For spin-${1\over2}    $ fermions, we shall focus  on the $2k_F$ susceptibilities for `site' $M=+$  and `bond' $M=-$  charge (CDW, BOW: \hbox{$\mu_p=0$}), and spin (SDW$_{x,y,z}$, BSDW$_{x,y,z}$: \hbox{$\mu_{p=1,2,3}$}) density-wave correlation of  the Peierls channel. 
The corresponding  composite fields are  
 \begin{eqnarray}
 {\cal O}^M_{\mu_p}(\omega,\Omega)  =  &{1\over 2} \, [O_{\mu_p}(\omega,\Omega) +M O_{\mu_p}^*(\omega,-\Omega)], &  \\  
  O_{\mu_p}(\omega,\Omega)=  \sqrt{T\over L}&\sum_{{k},\alpha\beta} \psi_{-,\alpha}(k-2k_F,\omega-\Omega)&\cr
  &\  \times \ \sigma_{\mu_p}^{\alpha\beta}\psi^*_{+,\beta}(k,\omega).& 
\end{eqnarray}
In the Cooper channel, we   consider the uniform superconducting singlet (SS: $\mu_c=0$) and triplet (TS$_{x,y,z}$: $\mu_c=1,2,3$) susceptibilities. The corresponding composite fields   at zero pair momentum are given by 
\begin{eqnarray}
{\cal O}_{\mu_c}(\omega,\Omega) &\!\!\! =\!\!\!\! & \sqrt{T\over L}\sum_{{k},\alpha\beta}\! \psi_{-,\alpha}(-k,-\omega+\!\Omega)\sigma_{\mu_c}^{\alpha\beta}\psi_{+,\beta}(k,\omega). \cr
  &  &   
\end{eqnarray}
For both channels, $\sigma_{0}={\bf 1}$ and \hbox{$\sigma_{1,2,3}=\sigma_{xyz}$} are the Pauli matrices.      

In the case of spinless fermions, only   the $2k_F$   `site' CDW    and BOW susceptibilities survive with
\begin{eqnarray}
\label{ }
 {\cal O}_{\mu_p}^M(\omega,\Omega)  &\!\!\! =\!\!\!\! &{1\over 2} \sqrt{T\over L} \sum_{k} [  \psi_{-}(k-2k_F,\omega-\Omega)\psi^*_{+}(k,\omega)\cr
 &+&M \psi^*_{+}(k+2k_F,\omega-\Omega)\psi^*_{-}(k,\omega))]. 
 \end{eqnarray}
In the superconducting channel,  only one susceptibility  is considered with the corresponding pair  field 
\begin{equation}
\label{ }
{\cal O}_{\mu_c}(\omega,\Omega) = \sqrt{T\over L}\sum_{{k}}\! \psi_{- }(-k,-\omega+\!\Omega) \psi_{+}(k,\omega).
\end{equation}

Adding (\ref{sources}) to the action $S$ in (\ref{Z}), the  partial integration (\ref{ZRG}) at the one-loop level yields additional outer shell contributions that correct $S_h$ and which gives  the recursion relation
\begin{eqnarray}
\label{Recurh}
    && S_h[\psi^*,\psi]_{\ell+d\ell} =    S_h[\psi^*,\psi]_{\ell}  + \langle \bar{S}_h\bar{S}_I\rangle_{\rm o.s}  + {1\over 2} \langle \bar{S}_h^2\rangle_{\rm o.s} +\ldots \cr
     &  &  
\end{eqnarray} 
The second term $\langle \bar{S}_h\bar{S}_I\rangle_{\rm o.s}$  is proportional to ${\cal O}_\mu h_\mu^{*}$ and its complex conjugate, and  leads to the flow equations for  the renormalization factors $z_\mu$ of the pair vertex parts.  In the   density-wave channel, its evaluation  leads to 
\begin{eqnarray}
 && \partial_\ell z_{\mu_p}^M(\omega,\omega+\Omega)=    \int_{-\infty}^{+\infty}  {d\omega'\over 2\pi} \Big\{z_{\mu_p}^M(\omega',\omega'+\Omega) \cr
&&\hskip 0.5 truecm \times \  \big[\, \tilde{g}_{\mu_p}(\omega,\omega',\omega+\Omega) + M\tilde{\cal G}_3(\omega,\omega'+\Omega,\omega')   \big] I_P(\omega',\Omega) \Big\},\cr
&&\ 
\label{ZPeierls}
\end{eqnarray}
where
\begin{eqnarray}
 \tilde{g}_{\mu_p=0}(\omega,\omega',\omega+\Omega) \!\!\!  &=&\!\!    \tilde{g}_2(\omega',\omega,\omega+\Omega), \cr
 && - 2 \tilde{g}_1(\omega,\omega',\omega+\Omega), \cr\cr
 \tilde{g}_{\mu_p\ne 0}(\omega,\omega',\omega+\Omega)\!\!\!  &=&\!\!\tilde{g}_2(\omega',\omega,\omega+\Omega), \cr\cr
 \tilde{\cal G}_3(\omega,\omega'+\Omega,\omega')  &=& \tilde{g}_3(\omega,\omega'+\Omega,\omega')  \cr  &&-2  \tilde{g}_3(\omega,\omega'+\Omega,\omega + \Omega), 
 \label{gZPS}
\end{eqnarray} 
for fermions with spins, and 
\begin{eqnarray}
\label{gZPS0}
 \tilde{g}_{\mu_p}(\omega,\omega',\omega+\Omega) & =& \tilde{g}_f(\omega,\omega',\omega+\Omega), \cr\cr
  \tilde{\cal G}_3(\omega,\omega'+\Omega,\omega')  &=&- \tilde{g}_3(\omega,\omega'+\Omega,\omega'),
\end{eqnarray}
in the  spinless case.
Similarly, for  the superconducting channel, one gets
\begin{eqnarray}
\label{ZC}
\partial_\ell z_{\mu_c}(\omega,-\omega+\Omega) & =  &  \int_{-\infty}^{+\infty}  {d\omega'\over 2\pi}\Big\{   z_{\mu_c}(\omega',-\omega' + \Omega)\cr
&  &\ \ \times \,\tilde{g}_{\mu_c}(\omega,\omega',\Omega) I_C(\omega',\Omega)\Big\},
\end{eqnarray}
where
\begin{eqnarray}
\label{gZC}
 \tilde{g}_{\mu_c=0}(\omega,\omega',\Omega)  & = & - \tilde{g}_1(\Omega-\omega,\omega,\omega')-  \tilde{g}_2(\omega,\Omega-\omega,\omega'), \cr\cr
 \tilde{g}_{\mu_c\ne 0}(\omega,\omega',\Omega)  & = & \, \tilde{g}_1(\Omega-\omega,\omega,\omega')-  \tilde{g}_2(\omega,\Omega-\omega,\omega').\cr
 && 
\end{eqnarray}
 for spin-${1\over 2}$ fermions,
and 
\begin{equation}
\label{gZCS0}
\tilde{g}_{\mu_c}(\omega,\omega', \Omega) = - \tilde{g}_f(\omega, \Omega-\omega, \omega')
\end{equation}
 in the spinless case.

As a result of the partial trace integration,  the last term  of (\ref{Recurh}), which is    proportional to $ h^{(M)*}_{\mu_{c(p)}}h^{(M)}_{\mu_{c(p)}} $,  is  generated along the flow and corresponds to the susceptibility in each channel considered, namely
\begin{eqnarray}
\partial_\ell \label{Chi}
 \chi_{\mu_{c(p)}}^{(M)}(\Omega)\!\!\! & = &(\pi v_F)^{-1}\!\!\! \!\int_{-\infty}^{+\infty} \! {d\omega\over 2\pi}\Big\{  |z_{\mu_{c(p)}}^{(M)}(\mp\omega,\omega+\Omega)|^2\cr
 && \times (2s+1)I_{C(P)}(\omega,\Omega)\Big\}, 
\end{eqnarray}
which  has been  defined positive ($\chi_{\mu_{c(p)}}^{(M)}(\Omega)=0$ at $\ell=0$), and where  $s$ is the spin.

We close this section by a digression on the numerical aspects associated with the solution of the above equations. Their numerical evaluation makes use of  patches in the frequency manifold. The frequency axis is discretized into a total of 15 subdivisions or patches  between the maximum values  $\omega_{\rm max}=\pm 1.5\:E_{F}$, which serve  as bounds of integration  for the  frequency. The interaction is taken as constant over each patch where  the loop integrals are    done exactly.  In order to reduce the number of  frequency dependent coupling constants, we take advantage of certain symmetries  namely,   the time inversion,  left-right Fermi points symmetry, and  the exchange symmetry between the   incoming $\left(\omega_{1}, \omega_{2}\right)$ and  outgoing $\left(\omega_{3}, \omega_{4}=\omega_{1}+\omega_{2}-\omega_{3}\right)$ frequencies. The last symmetry   antisymmetrizes   the initial conditions for the spinless fermions case, especially for the umklapp process $\tilde{g}_{3}$. We thus have to calculate 932 different functions   for each $\tilde{g}_{i}$. The same procedure is used to calculate the response functions and susceptibilities. The flow equations are numerically solved until the most singular susceptibility diverges with the slope $\pi v_F\partial_\ell \chi_\mu^M = 10^6$, which determines the critical value $\ell_c$ at which the algorithm is stopped.


\section{Results}
\subsection{Adiabatic  limit}
The results at non zero-phonon frequency will be compared to those of   the adiabatic limit where   $\omega_{0(D)}\to 0 $.  In this limit,  the initial conditions given  in Sec.~\ref{mod} for both  models show  that either $\omega_{3-1}\to 0$ or $\omega_{3-2}\to 0$, indicating that  no phonon exchange between fermions at finite frequency  is possible. In the spinless case the $g_f$ coupling Eq.~(\ref{gf}) reduces to its backward scattering part. Therefore only close loops contribute to the renormalization of both the coupling constants, susceptibilities, and  one-particle self-energy $\Sigma $; the latter being  vanishingly small in the adiabatic limit.

 The flow equations  for fermions with $s=1/2$ (resp.~$s=0$) (\ref{Flowg1}-\ref{Flowg3}) [resp. Eqs. (\ref{FlMCgfS0}-\ref{FlMCg3S0})]  can be recast into equations  for $g_1$ and $g_3$, which become independent of frequencies 
 \begin{equation}
\label{ }
\partial_\ell(\tilde{g}_1\pm \tilde{g}_3)   = -  (2s+1) (\tilde{g}_1\pm \tilde{g}_3)^2/2.
\end{equation}
 The solution is obtained at once 
 \begin{equation}
\label{g1g3}
\tilde{g}_1(\ell)\pm \tilde{g}_3(\ell) = {\tilde{g}_1 \pm \tilde{g}_3 \over 1+  {1\over 2}(2s+1) (\tilde{g}_1\pm \tilde{g}_3)\ell},
\end{equation} 
which presents a singularity at $\ell_0= -2[(2s+1)(\tilde{g}_1\pm \tilde{g}_3)]^{-1}$ for combinations of    bare  attractive couplings $\tilde{g}_1\pm \tilde{g}_3 $ found in  the MC (+) and SSH ($-$) models (Eqs. (\ref{IniMC})-\ref{IniSSH}), and (\ref{gf}-\ref{g3S0})). This signals an instability of the fermion system and the formation of a Peierls state with a $-$ mean-field (MF) $-$ gap   $\Delta_0 (\equiv 2E_Fe^{-\ell_0})$,   which takes the BCS form\cite{Caron84}
\begin{equation}
\label{MFGap}
\Delta_0 =2E_F \exp \Big(\!-2/ (2s+1)|\tilde{g}_1\pm \tilde{g}_3|\, \Big).
\end{equation}
This singularity is present in the pair vertex factors $z^M_{\mu_p=0}$ at $\Omega=0 $ in  either  CDW or BOW channel depending of the model. In the adiabatic limit this can be seen by retaining only closed  loops   in   (\ref{ZPeierls}), where for frequency independent couplings,  $z_{\mu_p=0}^M$ becomes in turn independent of $\omega$ and  obeys the following flow equation at $\Omega=0$
\begin{equation}
\label{ }
\partial_\ell \ln z_{\mu_p=0}^M = -  (2s+1) (\tilde{g}_1+M \tilde{g}_3)/2.
\end{equation}
With the help of  Eq.~(\ref{g1g3}), this is readily solved to lead  the simple pole expression  $ z_{\mu_p=0}^M=[1+  {1\over2}(2s+1) (\tilde{g}_1+M \tilde{g}_3)\ell]^{-1}$. From (\ref{Chi}), the $2k_F$ susceptibility   takes  the form
\begin{equation}
\label{MFChi}
\pi v_F\chi_{\mu_p=0}^M(\ell) = {\ell\over 1+  {1\over 2}(2s+1) (\tilde{g}_1+M\tilde{g}_3)\ell}.
\end{equation}
The expected simple pole divergence at $\ell_0$ then occurs   in   the   site CDW (\hbox{$M=+$)}  response for the MC  model and in  the  BOW(\hbox{$M=-$)} response  for the  SSH model. No enhancement is found for the susceptibilities in the superconducting channel. 

Strictly speaking, the above adiabatic MF  results hold for  models where only momentum independent couplings are retained. In the case of the SSH model for spinless fermions, however, the adiabatic limit of Eqs.~(\ref{gfSSHS0}-\ref{guSSHS0}) does not coincide with the MF result  due to the presence of $g_u$. In the adiabatic limit the flow equations read
\begin{eqnarray}
\label{AdiaSSH}
\partial_\ell \tilde{g}_1& = & -{1\over 2} \tilde{g}^2_1 -{1\over 2} (\tilde{g}_u+\tilde{g}_3)^2,\cr\cr
\partial_\ell \tilde{g}_3 & = & - \tilde{g}_3  \tilde{g}_1,\cr\cr
 \tilde{g}_u (\ell) &= &   \tilde{g}_u\exp\Big[-2\int_0^\ell (1+\tilde{g}_1(\ell))d\ell\Big].
\end{eqnarray}
The solution of these equations shows that the value of the  adiabatic  SSH gap $\Delta_0$  for spinless fermions  is slightly reduced   compared to the MF prediction (\ref{MFGap}) where $g_u$ is absent.

\subsection{The molecular crystal model }
\subsubsection{Spinless case  }
\begin{figure}
 \includegraphics[width=9.0cm]{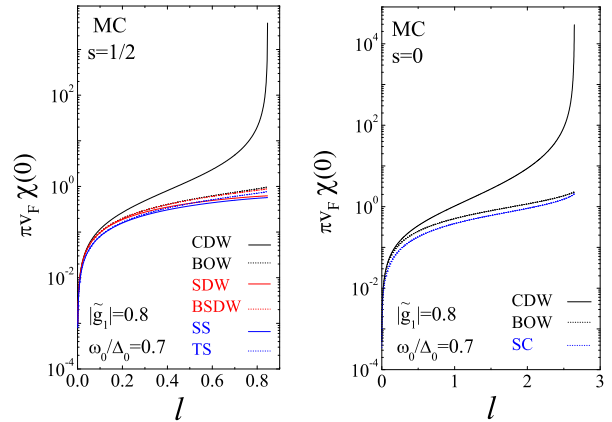}
\caption{\label{ChiMC} Typical variation  of the susceptibilities
with the scaling  parameter $\ell$ for the MC model for spinless  fermions($s=0$, right) and spin-${1\over 2}$ fermions  ($s=1/2$, left). The locus of the singularity at $\ell_c$ gives the value of the gap $\Delta=E_0(\ell_c)$.} 
\end{figure}

\begin{figure}
 \includegraphics[width=9.0cm]{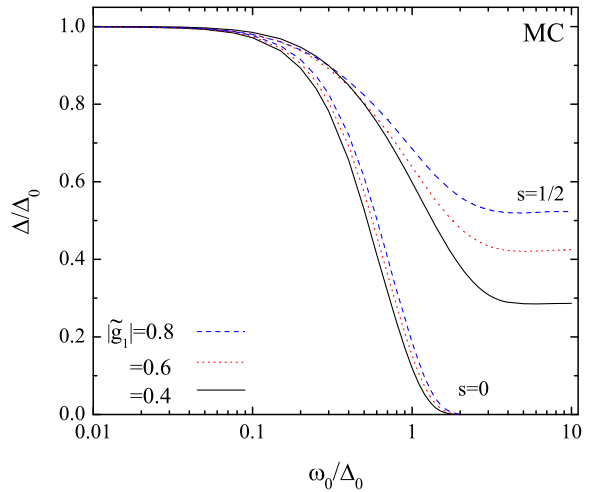}
\caption{\label{GapMC}  
 The site CDW gap of the MC model for  spinless ($s=0$) and spin-${1\over 2}$ ($s=1/2$)  fermions as a function of the phonon frequency and for different couplings.  Both quantities are normalized to the MF gap.} 
\end{figure}

The solution of the flow equations (\ref{FlMCgfS0}-\ref{FlMCg3S0}) for the MC model in the spinless  case ($s=0$) is obtained by using the antisymmetrized  boundary conditions given in  (\ref{gf}-\ref{g3S0}) at $\ell=0$.   The typical flow  of   susceptibilities  in the Peierls and Cooper  channels at an intermediate   phonon frequency is shown in Figure \ref{ChiMC}. Like for the MF result (\ref{MFChi}), the singularity  is found to occur solely in the site ($M=+$) CDW susceptibility at $\ell_c$. There is no noticeable enhancement of other responses including those of the superconducting channel. The singularity  signals  the existence of a Peierls gap $\Delta$ ($  \equiv2E_F e^{-\ell_c})$ with an amplitude that  is reduced at non zero  $\omega_0$ compared  to its  adiabatic value $\Delta_0$ (Eq. (\ref{MFGap})). Figure \ref{GapMC}  shows this renormalization  as a function of the ratio $\omega_0/\Delta_0$ of the phonon frequency to the MF gap (here the molecular mass $M_0$ is varied  while the spring constant $\kappa_0$ is kept fixed). For     small $\omega_0/\Delta_0$   the gap is weakly renormalized  and remains close to its classical value.  However, when the ratio $\omega_0/\Delta_0$ approaches unity,  the gap undergoes a rapid decrease due to quantum fluctuations. This results from the growing of vertex corrections  and interference between Peierls and Cooper scattering channels. These fluctuations signal    a change of regime (defined  at the point of a change of curvature for the gap profile)   that we refer to as a quantum-classical crossover for the gap .

\begin{figure}
 \includegraphics[width=9.0cm]{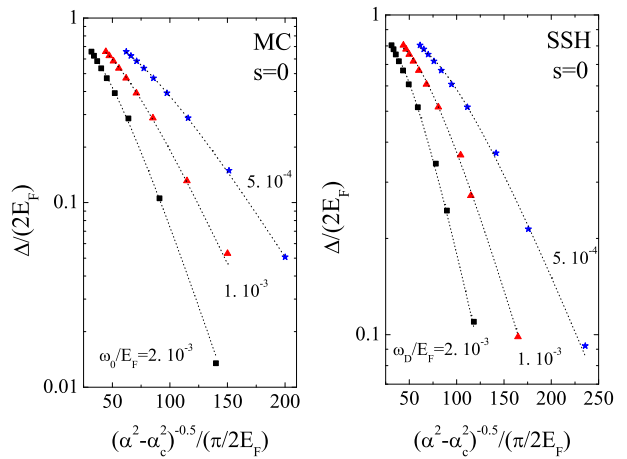}
\caption{\label{GapBaxter}  
 The site CDW (left) and BOW (right)  gaps  as a function of the critical parameter of the KT  transition to  a Luttinger liquid for the CM and SSH models in the spinless case. The dotted line is a least squares  fit to the Baxter formula Eq.~(\ref{Baxter}).} 
\end{figure}

The remaining gap tail terminates with   a transition to a $\Delta =0$ disordered state  at a threshold  frequency   slightly above $\Delta_0$. The   ratio $\omega_0/\Delta_0$   at which the transition occurs  is   weakly    dependent on the initial   $\tilde{g}_i$ for the range of coupling covered by the present RG. This  result   corroborates  the old two cut-off scaling arguments for the disappearance of an ordered state at $\omega_0\sim \Delta_0$,\cite{Caron84} and agrees with  the DMRG, \cite{Bursill98}  and   Monte Carlo\cite{Hirsch83} results for the MC model. The nature of  the transition  to  the quantum  gapless   state is also of interest. We follow the notation of Ref.\cite{Caron96} and   define the  coupling $\alpha    \equiv {1\over 2}\sqrt{|\tilde{g}_1|\omega_0E_F}.  $ We see from Figure \ref{GapBaxter} that the variation of the gap $\Delta$, close to the critical $\alpha_c$ at which the transition occurs,  follows closely the  Baxter formula for a Kosterlitz-Thouless (KT) transition \cite{Baxter82}
\begin{equation}
\label{Baxter}
\Delta  \propto  {2E_F\over\sqrt{\alpha^2-\alpha^2_c}}\, e^{-b/\sqrt{\alpha^2-\alpha^2_c}},
\end{equation} 
where $b$ is positive a constant. This    behaviour found   in the weak coupling range  is similar to the one obtained by the DMRG method and perturbative expansion in  strong coupling. \cite{Bursill98,Hirsch83} 

\begin{figure}
\includegraphics[width=9.0cm]{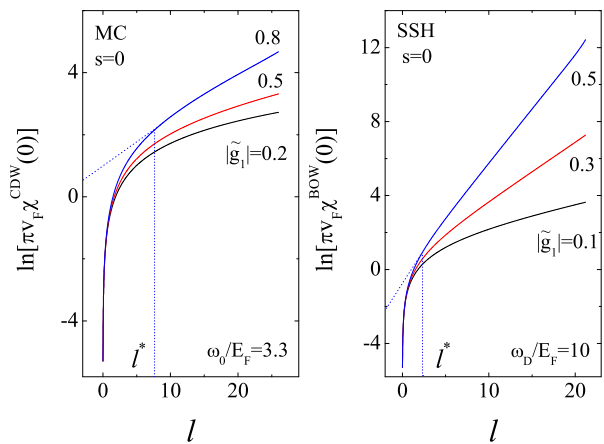}
\caption{\label{chiLL}  
Typical power law divergence of the site CDW (left)  and BOW (right) susceptibilities at   $\ell\gg \ell^*$ in the gapless Luttinger liquid regime.  } 
\end{figure}

\begin{figure}
\includegraphics[width=9.0cm]{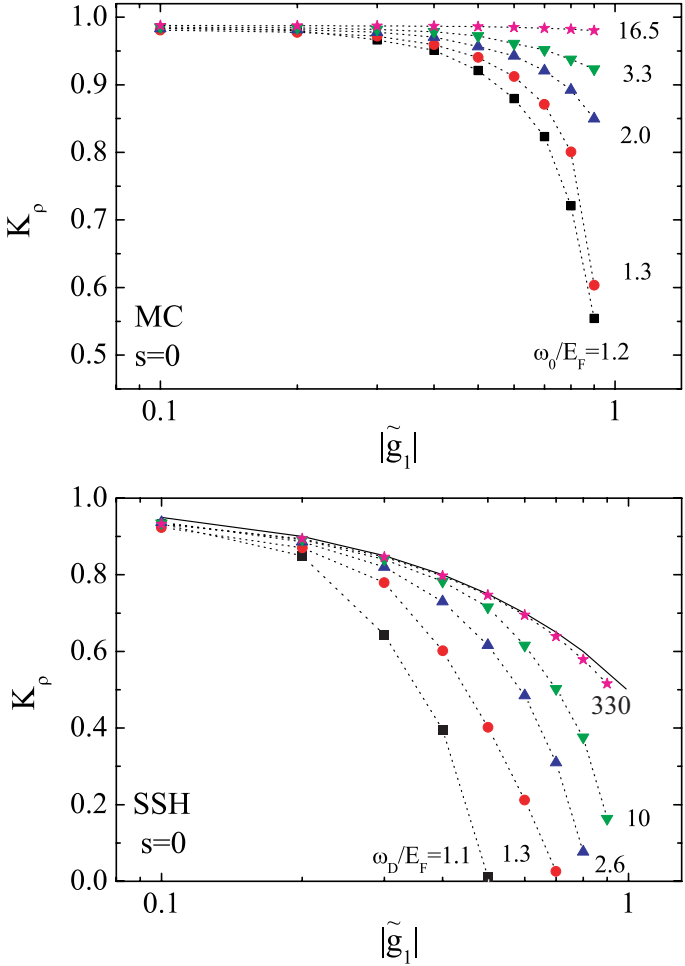}
\caption{\label{KrhoCMSSH}  
One-loop calculation of the density stiffness parameter $K_\rho $  of  the MC (upper panel) and SSH (lower panel) models in the spinless case as a function of the initial coupling $|\tilde{g}_{1}|$, and for different phonon frequencies. The continuous line in the lower panel corresponds to the antiadiabatic one-loop result $K_\rho=1-{1\over 2} |\tilde{g}_1| $.} 
\end{figure}

For phonon frequency above the  threshold,  the  new state is expected to be a Luttinger liquid.\cite{Caron84,Bursill98} This is seen at the  one-loop RG level from  the existence of a power law  behaviour of the site CDW susceptibility, namely $ \chi^+_{\mu_p=0}(\ell) \sim [E_0(\ell)]^{-\gamma }$. The latter   takes place only above some characteristic $\ell^*$  (Fig.~\ref{chiLL}) that  depends on   $\omega_0$ and  which decreases with the strength of the coupling.  Non universality   is also found for the Luttinger liquid exponent $\gamma$  for  $E_0(\ell)\ll E_0(\ell^*)$.  Following the   one-dimensional theory,\cite{Giamarchi04,Shankar90}   the exponent can be written as $\gamma= 2-2K_\rho$, where  $K_\rho$ is the   stiffness parameter for  the density degrees of freedom   that enters in the bosonization scheme.  Within the limitation of a weak coupling theory, it  is therefore possible to determine the dependence of $K_\rho$  on interaction and  phonon frequency.   As   shown in  Figure \ref{KrhoCMSSH}, the one-loop RG results   confirms  the non universal  character   of the stiffness parameter. Going  down on the frequency scale,       $K_\rho$ is sizeably  reduced at the approach of the KT transition   where retardation effects   have a strong influence on the properties of  the Luttinger liquid parameter. We find that  $K_\rho$ stays above the minimum value of ${1\over 2}$ known for  isotropic  spin chain  in the  gapless domain\cite{Haldane82,Shankar90} -- following the Wigner-Jordan transformation of    spins into  spinless fermions.   On the other hand, $K_\rho $ is only weakly dependent on the couplings  at large $\omega_0$,  where it tends to      the   non adiabatic -- coupling independent --  value $K_\rho= 1$ at $\omega_0 \to \infty$.    Recall that the initial couplings of the MC model  (Eqs.~(\ref{gf}) and (\ref{g3local}))    vanish in this limit, and the system  is equivalent to a non interacting Fermi gas.

\begin{figure}
 \includegraphics[width=9.0cm]{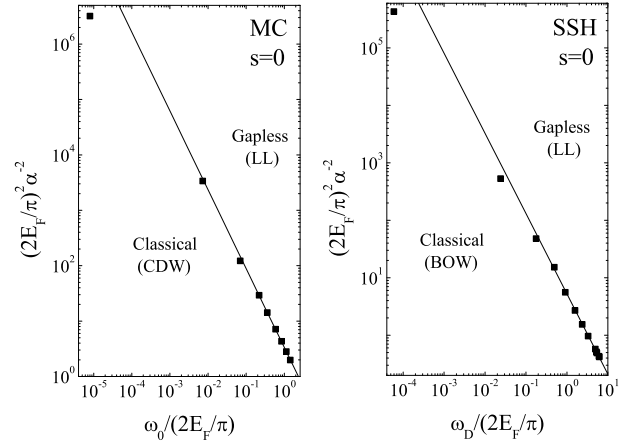}\caption{\label{MCSSHphase}  
Phase diagram of MC (left) and SSH (right)  models for spinless fermions. The full squares are the RG results and the continuous   lines is  the power law $\alpha^2_c\sim \omega_0^{1.4}$  of the critical coupling of the KT transition  on  the phonon frequency.    } 
\end{figure}

From the variation  of the critical coupling $\alpha_c$ with   the phonon frequency $\omega_0$, one can construct the phase diagram  of Fig.~\ref{MCSSHphase}. The phase boundary between the  insulating and the metallic Luttinger liquid  states is found to follow closely a power law dependence of the form $\alpha_c\sim \omega_0^\eta$, with the exponent $\eta\approx0.7$.  This   feature captured by a one-loop calculation is analogous to the quantum-classical  boundary of the phase diagram of the 1D  XY spin-Peierls model   determined by the DMRG   method.\cite{Caron96}  The latter model is  also characterized by a zero temperature KT  transition as we will see for the spinless SSH model in Sec.~\ref{SSSH}. 

\subsubsection{Spin-${1\over 2}$ fermions}
The results for the MC model with  spin-${1\over 2}$ fermions   ($s=1/2 $) ensue from the solution of Eqs.~(\ref{Flowg1}-\ref{Flowg3}) and the computation of the susceptibilities (\ref{Chi}), from (\ref{ZPeierls}-\ref{gZPS}) and (\ref{ZC}-\ref{gZC}). Like spinless fermions, the singularity   is  found in the $M=+$,  site CDW susceptibility at   finite $\ell_c$ (Fig.~\ref{ChiMC}).  The corresponding value for the gap $\Delta$ is  reduced with respect to the adiabatic mean-field result $\Delta_0$ in (\ref{MFGap}). The onset of   quantum fluctuations due to growing interference between different scattering channels is again responsible  for a quantum-classical crossover  when   $\omega_0$ appoaches $  \Delta_0$ where there is change of curvature in the gap profile, but the gap never goes to zero. It remains finite at  large phonon frequencies and is dependent on the bare attractive amplitude  $\tilde{g}_i$. At large frequency  the singularity at $\ell_c$ occurs essentially independently for   spin  [$\tilde{g}_1(\{\omega\})$] and charge [$2\tilde{g}_2(\{\omega\})-\tilde{g}_1(\{\omega\})$,  $\tilde{g}_3(\{\omega\})$] combinations of couplings at zero   Peierls  and Copper frequency.    As a function of  $\omega_0$, the system then undergoes a crossover from a renormalized classical Peierls state towards  a quantum but still site-CDW ordered state in which both spin and charge degrees of freedom are gapped due to attractive couplings and the relevance of umklapp processes at arbitrary large but finite $\omega_0$.       An ordered state is well known to be found at large $\omega_0$  in Monte carlo simulations.  \cite{Hirsch83}   This quantum-classical crossover marks the onset of a decoupling between spin and charge degrees of freedom, a separation found   in the Luther-Emery model.\cite{Luther74,Emery79,Giamarchi04} 

 It is worth noting that in the purely non adiabatic case where $\omega_0$ is strictly  infinite,  the initial couplings (\ref{IniMC}) are independent  of frequency and satisfy  the conditions   $\tilde{g}_1<0$, and $\tilde{g}_1-2 \tilde{g}_2=|\tilde{g}_3|$, which  coincide with those of an attractive Hubbard model. Its exact solution is well known to give a disordered ground state. At the one-loop level, the  RG equations  (\ref{Flowg1}-\ref{Flowg3}) at zero external frequencies show indeed   that    $\tilde{g}_1$ alone  is singular, with a gap  in the spin sector only. Umklapp processes are irrelevant and   charge degrees of freedom remain  gapless,  consistently  with  the absence of   long-range order   at $\omega_0=\infty$. \cite{Dzyaloshinskii72,Kimura75,Solyom79,Emery79}   Working at arbitrarily large but  finite $\omega_0$ introduces   finite retardation effect that is sufficient to make initial conditions deviate from those of the  attractive Hubbard model. This   restores the relevance of umklapp term in the charge sector and  in turn long-range order.\cite{Hirsch83}

\subsection{\label{SSSH}SSH}

\subsubsection{Spinless case}
\begin{figure}
 \includegraphics[width=9.0cm]{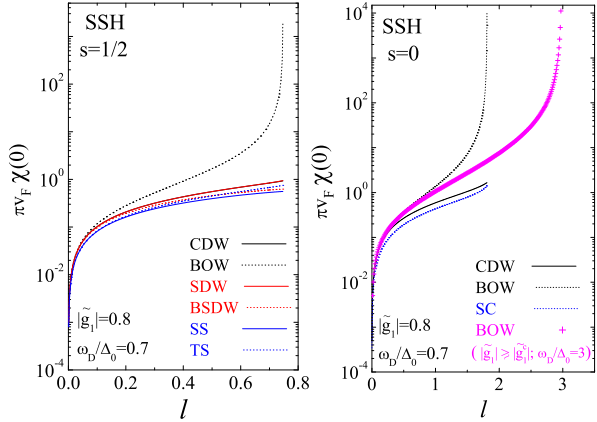}
\caption{\label{ChiSSH} Typical variation  
of the susceptibilities for the SSH model as a function of the scaling parameter $\ell$ for spinless  fermions ($s=0$, right) and spin-${1\over 2}$ fermions with spins ($s=1/2$, left).} 
\end{figure}

We turn now to the study of the SSH model. In the spinless case the presence of the non local umklapp term $g_u$ introduces some qualitative differences  with  the  MC model  for which  this term is absent. Thus the solution of (\ref{gfSSHS0}-\ref{guSSHS0}), and  (\ref{ZPeierls},\ref{ZC},\ref{Chi}) in the spinless case shows that  for small $\omega_D/\Delta_0$,  the BOW   susceptibility  (\hbox{$\mu_{p=0}, M= -$}) is the only singular response  that leads to   a gap  at zero temperature  (Fig.~\ref{ChiSSH}). 

\begin{figure}
 \includegraphics[width=9.0cm]{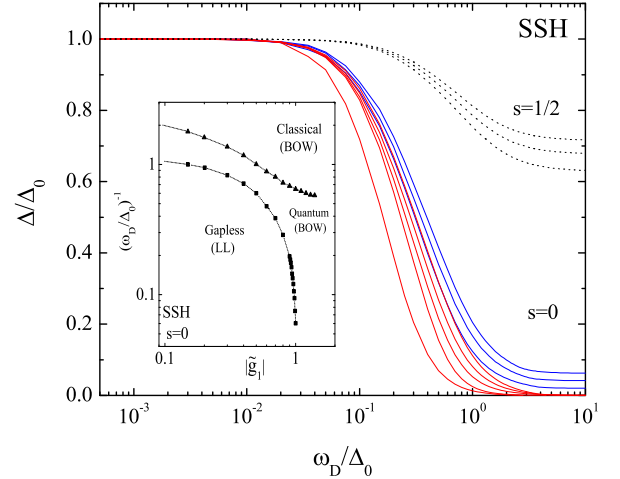}
\caption{   \label{SSHgap}
The BOW  gap normalized to its adiabatic value as a function  the ratio of the phonon frequency and the adiabatic gap  for the SSH model for spinless fermions  (continuous lines, left to right, red: $|\tilde{g}_1|= 0.1, 0.4, 0.6, 0.9 < |\tilde{g}^c_1|$;  blue: $|\tilde{g}_1|= 1.1, 1.15, 1.2 > |\tilde{g}^c_1|)$) and  spin-${1\over2}$ fermions (doted lines; from bottom to the top: $|\tilde{g}_1|=0.5, 0.7, 0.8)$. Inset: phase diagram of the spinless SSH model.  Quantum BOW order/gapless  Luttinger liquid (LL)  (squares), and the quantum-classical crossover  for the  BOW  order state (triangles).   } 
\end{figure}

As one moves along the $\omega_D/\Delta_0$ axis (along which the mass $M_D$ varies and $\kappa_D$ is   constant),  one finds again  from Fig.~\ref{SSHgap}, that for $\omega_D/\Delta_0<0.1$, the gap is weakly renormalized compared to its  adiabatic classical value computed from (\ref{AdiaSSH}). As  the ratio  increases further,    there is a strong downward renormalization for the gap which undergoes a quantum-classical crossover. However, at variance with the MC model,  the ratio $ \omega_D/\Delta_0$ at which there is a change of curvature in the variation of the  gap shows a  stronger dependence on the amplitude of  the initial coupling   --  parametrized by the backward scattering part $|\tilde{g}_1|$ of $\tilde{g}_f$ in (\ref{gf}) (triangles,  inset of Fig.~\ref{SSHgap}).

  At  higher  phonon frequency we come up against a critical  value   where the gap completely vanishes and the system enters in a metallic state at zero temperature. This critical ratio is also coupling dependent (squares, inset of Fig.~\ref{SSHgap}).  In  the non adiabatic limit when   $(\omega_D/\Delta_0)^{-1}\to 0 $,
 the critical coupling  heads on to   the one-loop limiting value  $ |\tilde{g}^c_{1}|=1$,  which can  be extracted directly  from  Eq.~(\ref{guSSHS0}) in this limit. A frequency dependent threshold  $|\tilde{g}^c_1|$  for the existence of an ordered state is a direct consequence of  the relevance of the non local umklapp term $\tilde{g}_u$ in (\ref{guSSHS0}), which differs markedly from the MC model and simple  two-cutoff scaling arguments especially at large phonon frequency.   The  phase diagram shown in  the  inset of Fig.~\ref{SSHgap} can then be obtained for the  spinless SSH model. A critical line for $(\omega_D/\Delta_0)^{-1} $  {\it vs} $|\tilde{g}_1^c|$  can    be drawn, separating the quantum   disordered state  from   BOW order. The singularity of the BOW susceptibility in the quantum   domain  is shown in Fig.~\ref{ChiSSH}.  
 
  The numerical solution of the flow equations is not   carried out easily in the limit of very small couplings owing to  the large number of frequencies needed to reach the desired accuracy. Our results, obtained down to $|\tilde{g}_1|=0.1$,   tend  to show, however, that the critical line $|\tilde{g}^c_1 | $  extrapolates to zero   at the finite value of the ratio $\Delta_0/\omega_D  \approx 1.1 $,  which joins the  value   obtained   from the two-cutoff scaling arguments.\cite{Caron84} Above this value,  an  ordered BOW  state would then  be  found at  any finite coupling. When the ratio finally crosses  the quantum-classical line, the system  enters in  a   Peierls BOW state similar to the one of  the classical adiabatic limit.  This quantum-classical boundary,  clearly identified in  Fig.~\ref{SSHgap} as a change of regime for the gap, is consistent with the one found by DMRG for the XY spin-Peierls chain (following the conversion of spins into spinless Wigner-Jordan fermions).\cite{Caron96}  

As one moves from the quantum massive domain towards the critical line at higher frequency, the gap collapses to zero.  Following the example of what has been done for the MC model, we follow the notation of Ref. \cite{Caron96} and define  $\alpha_c    \equiv {1\over 2}\sqrt{|\tilde{g}^c_1|\omega_0E_F},  $  as the critical coupling   where the gap vanishes.  We thus  find that close to the transition, $\Delta$   decreases to zero according   to the Baxter expression Eq.~(\ref{Baxter}) for a   KT  transition (Fig.~\ref{GapBaxter}). This result which carries over the whole critical line at finite frequency  is in accord with DMRG results obtained on the spin-Peierls XY  \cite{Caron96} and   XXZ\cite{Bursill99}  chains. For the latter model a KT transition was also found by  Citro {\it et al.,} \cite{Citro05} using  the RG method in the bosonization framework. In the same vein, Kuboki and Fukuyama also show by perturbation theory  that retardation is equivalent at large frequency to frustration  in the spin interactions,\cite{Kuboki87}  which beyond some threshold is well known to promote a KT  transition to a dimerized state. \cite{Haldane82}

  As shown by the phase diagram of  Fig.~\ref{MCSSHphase}, the critical coupling   is found to follow  the power law variation  $\alpha_c\sim \omega_D^\eta$, with $\eta\approx 0.7$, over a sizeable range of the phonon frequency (deviations are found in the limit of small frequency). Such a power law  agrees with  the one found  in DMRG for the XY spin-Peierls chain,\cite{Caron96} and  is similar to the   one obtained in  the MC case  for  spinless fermions.

As regards to the nature of the gapless  liquid state in the disordered region, the situation is qualitatively similar to the MC model. We find the presence of a non universal  power law divergence for  the BOW response  function $\chi_{\mu_p=0}^-(2k_F) \propto [E_0(\ell)]^{-\gamma}$  below some characteristic energy scale  $E_0(\ell^*)$ (Fig.~\ref{chiLL}), which indicates the presence of  a Luttinger liquid. The  weak coupling determination of the charge  stiffness parameter $K_\rho (\gamma = 2-2K_\rho)$\cite{Giamarchi04} at the one-loop level  is shown in Figure \ref{KrhoCMSSH}. By comparison with the MC model, $K_\rho$ is smaller and shows a stronger variation with the strength of the coupling even for large phonon frequencies. This is so for in the   non adiabatic limit, the initial conditions for both  $\tilde{g}_f(\{\omega\})$ and $\tilde{g}_u(\{\omega\})$ are non zero and  a massive phase remains possible. In this limit, the RG  results join  the   one-loop relation $K_\rho=1-{1\over 2} |\tilde{g}_1| $ (continuous line, lower panel  of Fig.~\ref{KrhoCMSSH}),  which is known to be a good approximation to the exact result.\cite{Schulz00,Baxter82}  It is worth noticing that although the transition remains of infinite order for any path that crosses the critical line from the massive sector to the Luttinger liquid one,  the characteristics of the latter phase, through its exponent $\gamma$ (or its stiffness coefficient $K_\rho$) is strongly dependent on retardation effect.  With the caveat of the limited accuracy of one-loop calculations for sizeable $\gamma$,  the present results  would indicate that except for the domain of large frequency, $K_\rho$ penetrates deeply into  the `Ising  sector' where $K_\rho < 1/2$   at the approach of the critical line.  

A similar downward renormalization of $K_\rho$ by retardation effects has also been found by Citro {\it et al.,} \cite{Citro05} using the self-consistent harmonic approximation and the RG method  in the bosonization frame  for the XXZ spin model of the spin Peierls instability.  When the  spins are converted into fermions  through a Wigner-Jordan transformation, the properties of this model  are encompassed by the flow equations (\ref{gfSSHS0}-\ref{guSSHS0})  following a redefinition of the initial conditions (\ref{gf}-\ref{guk}).

 \subsubsection{ Spin-${1\over 2}$ fermions }
The results  for  the SSH model with spin-${1\over 2}$ fermions  is obtained from  the solution of Eqs.~(\ref{Flowg1}-\ref{Flowg3}) using the initial conditions (\ref{IniSSH}).  The computation of the susceptibilities (\ref{Chi}) using  (\ref{ZPeierls}-\ref{gZPS}) and (\ref{ZC}-\ref{gZC}) shows that the singularity and  the formation of the gap remains  as expected in the BOW channel (Fig.~\ref{ChiSSH}).  As a result of   growing   interference between the Peierls and Cooper channels  and vertex corrections in the scattering amplitudes,    the reduction of the BOW classical gap as a function of the frequency,  (Fig.~\ref{SSHgap})  is less pronounced than for the MC model. This reduction then evolves  to  a quantum-classical crossover at   $\omega_0\sim \Delta_0 $.  Following the example of the MC model, the system remains   massive for both spin and charge and is thus  BOW ordered in the quantum regime. The amplitude of the gap at large frequency is however bigger.   As a matter of fact, in the non adiabatic case where $\omega_D$ is infinite, the initial couplings (\ref{IniSSH}) are frequency independent but  at variance with the MC model, they   satisfy the inequalities $\tilde{g}_1<0$, $\tilde{g}_1< \tilde{g}_3 $.   These are compatible with the  Luther-Emery conditions for a mass in  both spin and charge sectors. \cite{Luther74,Emery76b,Emery79} 
\section{Conclusion}
In this work we used an extension of  the RG approach to one-dimensional fermion gas that includes the full influence of retardation   in the interactions induced by phonons.    Within the inherent  bounds of  a weak coupling 
theory,  the method  has been put to the test and proved to be rather satisfying,  providing  a continuous description  of the gap as a function of  the  phonon frequency for  electron-phonon models  with either spinless or spin-${1\over 2}$ fermions.  Generally speaking, the results brought out the importance  of the static  scale $\Delta_0$ of the adiabatic theory for the occurrence of a quantum-classical crossover  for the gap as one cranks up  the phonon frequency, confirming in passing   the old  arguments of the two-cutoff scaling approach.   The RG calculations allowed us  to study  the nature of the transition to the gapless liquid phase for spinless fermions.  For both  the MC and SSH models,  this transition was found to be of   infinite order, consistently with existing numerical results.  

The RG method for  the SSH model  required   to take into account the momentum dependent umklapp term. The latter is responsible for  the continuation of the  infinite order critical line to arbitrary  large phonon frequency  where it connects to the well known results of frustrated spin chains. The existence in the   gapless phase of a sharp power law behaviour of $2k_F $ density response at low energy showed that this phase can be  identified with   a Luttinger liquid. The non universal variation of the  power law exponent with the strength of interaction and retardation was  obtained at the one-loop level.  Retardation effects induce  a downward renormalization of the Luttinger  parameter $K_\rho $  for both models in the disordered phase. However, this renormalization is apparently much stronger   in the SSH case  where  $K_\rho $ goes under  its limiting ${1\over 2}$ value known for the anisotropic spin chain in the gapless regime.

While  this work   did not dwell  on the combined influence of direct    and retarded interactions on Peierls-type instabilities, Coulomb  interaction can be actually incorporated   without difficulty  following  a mere change of the boundary conditions for the RG flow of  scattering amplitudes. A further extension of the method   that includes both the frequency and momentum functional dependence of the scattering amplitudes would be also worth while. As was shown very recently by Tam {\it et al.,}\cite{Bakrim1} in the context of the one-dimensional MC-Hubbard model and by Honerkamp {\it et al.,}\cite{Bakrim1} in the two-dimensional situation, 
the difficulties inherent to such an extension proved not insurmountable. 
An RG implementation of this sort for    interacting  quasi-one-dimensional  electron systems would   be quite desirable. It would  yield    a  more complete  description of electronic phases found in correlated materials like the organic conductors and superconductors.   The coupling of electrons to both intramolecular and intermolecular (acoustic) phonon modes are   in practice both present in these systems and their  characteristic energies are often close to the energy scales associated to the various types
of long-range order observed.\cite{Bourbon99} These systems then fall  in the intermediate  phonon frequency range considered  in this work, and for which retardation effects may play a   role in the  structure of their phase diagram.      The impact of retarded  interactions on   electronic states found in quasi-one-dimensional conductors  is currently under  investigation.  
\begin{acknowledgments}
We thank    L.~G. Caron  and K.-M. Tam
 for  useful discussions and comments.  H. B. thanks I. A. Bindloss for an helpful correspondence.   C.~B. also  thanks the Natural
Sciences and
Engineering Research Council of Canada (NSERC),  and the
 Canadian Institute for Advanced  Research   (CIFAR)  for
financial support.  \end{acknowledgments}

\end{document}